# Controlled Film Flow in Granulation of Metals for the Development of Amorphous Superhard and Functionally Unique New Materials

*Ivan V. Kazachkov*

Division of Information Technology and Data Analysis,
Nizhyn Gogol state university, 16600, Ukraine;
Division of Heat and Power, Royal Institute of Technology
Sweden, Ivan.Kazachkov@energy.kth.se

**Abstract.** The problem of granulation is very bright by the granulated materials, as well as by their application. In the paper, some history of the granulation problem during over century and modern applications of the metallic granulates and amorphous materials are given at the beginning. Then the specific own granulation problem is presented, which has concern to the controlled liquid metal jet and film flows for a production of the uniform by size and form particles (granules) cooled with a high rate, to be amorphous or close to the amorphous materials. Such granules of the given size and form are needed for the new material science. The basics of developed theory of the controlled jet and film flow disintegration with further rapid cooling of the drops obtained after flow disintegration are presented together with the new patented granulation devices. The developed methods and devices can be used for production of the amorphous or close to amorphous granules in a wide range of the given sizes, with very narrow (plus-minus 50% deviation of size from the average one).
**Keywords**: film flow disintegration, drops, granulation, amorphous material, high cooling rate

## 1. Introduction to the problem

Granulation of the diverse materials is widely used in a number of different applications such as pharmacy, food, and agriculture (granulated fertilizers), chemical technology, metallurgy, etc. In many industries, the granulation represents an important preliminary stage in a manufacturing of a product. In some cases, the granules are intended for a direct sale as a final product too. The reasons for transforming the powders or dust into granules can be an improvement of the flow properties (agglomeration of the filter dust), an increase of the bulk density (agglomeration of microsilica), and avoidance of the demixing effects (the ceramic press bodies). Development of the amorphous metallic granules is one of the perspective directions connected with a production of the new (including nanostructured) functionally unique materials.

An excellent ratio of a weight to strength, together with an appearance of the high-quality surface, made the ***magnesium alloys*** particularly suitable for the mobile applications in the consumer goods sector. In automotive applications, the magnesium die castings are increasingly replacing steel- and aluminium-based components. This is a major factor of the reduction in a total weight of the vehicles representing a valuable contribution to achieving the agreed climate targets. An improved the weight/strength ratio in a comparison to the aluminium alloys has some new applications as well: the aviation, aerospace and the other industries. Also, the magnesium granules (coated and uncoated, size of particles over about 160 μm) are used to reduce sulfur in the pig iron. The sequential process with lime and/or calcium carbide as the desulphurization reagents uses the both, pure magnesium and Mg mixtures. They ensure a highly efficient reduction of the sulfur to the lowest ratios in a short time of the running processes. Also, the special passivated Mg granules are used for the production of a cored wire, in a treatment of the nodular cast iron, and the hot metal desulphurisation. The magnesium powder (size of the particles less than about 160 μm) is used in pyrotechnics and for military purposes, the aluminium-magnesium granules are needed for space industry, etc.

## 2. Short history of granulation development and application

Probably the first attempts to use the granulation technology for improvement of the metals were done after the Second World War by Forton Harold R. (Method and apparatus for forming a powder from metals, 1945), Polysius Gmbh (Vorrichtung zum Granulieren von Pulver- oder Griessfoermigem Gut, 1952), Western Electric Co (Apparatus for making metal pellets, 1953), and United States Steel Corp (Disc-type balling device, 1955). Hundreds of patents on the subject appeared starting since the 1960-th from dozens of different companies and engineers. The first were Ibm (Amorphous alloys and process therefor, 1964), Matsushita Electric Ind Co Ltd (Ferromagnetic materials, Magnetic permeability material, 1966), United Aircraft Corp (Process for making filamentary materials, 1966 and Electrostatic coatings, 1967), Monsanto Co (Method for forming fibers and filaments directly from melts of low viscosities, 1969), Uddeholms Ab (Method of making granulate, 1970).

Metglas, Inc. (then Allied Signal, NJ) pioneered (1970s) the development and production of the amorphous metal, a unique structure alloy, in which the metal atoms are randomly distributed. The key to the company's proprietary manufacturing process is the rapid solidification of the molten alloy at a rate of approximately 1 mln Celsius degrees a second. The amorphous metals have a unique non-crystalline structure. Their excellent physical and magnetic properties combine the strength and hardness with the flexibility and toughness. These products help companies in reducing the operating costs, as well as in strengthening the energy conservation efforts and increasing the application efficiency. Then it was continued by a number of companies: Fuji Photo Film Co., Ltd (Ferromagnetic metal powder comprising lead and method for making the same, 1974), Allied Chemical Corporation (Method of producing amorphous cutting blades, Titanium-beryllium amorphous alloys, 1975; Metallic glasses with high crystallization temperatures and high hardness, 1976), The Research Institute for iron, steel and other metals of the Tohoku University (Iron-chromium series amorphous alloys, 1975), International Business Machines Corporation (Method for inducing uniaxial magnetic anisotropy in an amorphous ferromagnetic alloy, 1974; Controlled catalyst for manufacturing magnetic alloy particles with selective coercivity, 1975), University of Pennsylvania (Method of making amorphous metallic alloys with enhanced magnetic properties using the tensile stress, 1976), Bell Telephone Laboratories, Inc. (Electric fuse, 1976).

Afterward the above-mentioned companies continued their work in this exciting direction and many other joined to: General Electric Company, United Technologies Corporation, Kokusai Westinghouse Electric Corp.Denshin Denwa Kabushiki Kaisha, Western Gold And Platinum Company, Siemens Aktiengesellschaft, Tokyo Shibaura Denki Kabushiki Kaisha, Electric Power Research Institute, Inc., Tsuyoshi Masumoto, etc. Recently the following efforts were made: Glassimetal Technology, Inc. (Bulk iron-nickel glasses bearing phosphorus-boron and germanium, 2013), Yale University (Bulk metallic glass nanowires for use in energy conversion and storage devices, 2011), Hon Hai Precision Industry Co., Ltd. (Coated article and method for making the same, 2011), Lawrence Livermore National Security, Llc (Amorphous Metal Formulations and Structured Coatings for Corrosion and Wear Resistance, 2011), Glassimental Technology, Inc. (Bulk nickel-silicon-boron glasses bearing iron, 2014).

## 3. The most exciting accomplishments and challenges in granulation and applications

For the moment, there are known hundreds of publications and practical achievements reported from different companies in a field of granulation, amorphous materials, and different products of them. Let us just shortly mention some of the known results in this field [1-38]. The novel different metal alloy compositions in the amorphous state have been obtained in the world, in the USA, UK, Japan, India, Russia, Ukraine, and other countries as well. These new materials were superior to such previously known alloys based on the same metals. The new compositions were quenched to the amorphous state due which they obtained the desirable unique physical properties. For example, a novel of wire of these novel amorphous metal alloys was disclosed, as well as of other compositions of the same type. A limited number the amorphous (noncrystalline, glassy) metal alloys have been prepared by a rapid quenching of a molten alloy of a suitable composition. Alternatively, a deposition technique may be used when a suitably employed vapor is deposited, sputtered, electrodeposited, or chemically deposited to produce the amorphous metal. Production of the amorphous metal by the known techniques, either through a rapid quench of the melt or through a deposition, severely limits the form in which the amorphous metal is obtained. When the amorphous metal is obtained from the melt, the rapid quench is generally achieved by spreading of the molten alloy in a thin layer against a metal substrate (Cu or Al) held at the room temperature or below it. Typically, the molten metal is spread to a thickness of about 0.005 cm, which leads to a cooling rate of about $10^6$ C/s [19, 20].

Various methods have been invented for rapid quenching by spreading the molten liquid in a thin layer against a metal substrate [2-5]. For example, a gaseous shock wave propels a drop of molten metal against a substrate made of a metal such as copper [2], the piston and anvil technique [3], in which two metal plates come together rapidly and flatten out and quench a drop of molten metal falling between them; the casting technique [4] when a molten metal stream impinges on the inner surface of a rapidly rotating hollow cylinder open at one end; and the method of rotating double rolls [5], in which the molten metal is squirted into the nip of a pair of rapidly rotating metal rollers. These methods produce small foils or ribbon-shaped samples with one dimension much smaller than the other two, therefore they are not so useful in a practical matter because some important applications are severely limited due to big difference in properties of the material obtained in one of the directions (the layered material produced has some weak properties due to this).

Since the cooling rates necessary to obtain an amorphous state from quenched liquid metals and alloys are generally high or very high, amorphous metals are not formed in a form that does exclude sufficient cooling. Typically, they must have at least one size, small enough to allow a sufficiently rapid removal of heat from the solidifying material. Later we describe how we solved this problem. To avoid crystallization, the cooling rate necessary to achieve a stable amorphous state depends on the composition of the alloy. Some of the alloys can be obtained in an amorphous state with a lower cooling rate, which in practice can be more easily obtained or can be obtained with a greater thickness during melt quenching. There is a small set of compositions, for each of which it is easy to get an amorphous state. As before, the problem arises of obtaining an amorphous metal with predetermined processing conditions. And which alloy is the best glass-forming agent; this is a question for research.

Amorphous and crystalline states differ in the corresponding periodicity of the long-range or its absence. Compositional ordering in alloys is probably different for two states because of differences in their diffraction behavior of x-rays. Their measurements most often distinguish between crystalline and amorphous substance. Amorphous substance exhibits a slowly varying diffracted intensity similar to liquid, while crystalline materials give a much more rapidly varying diffracted intensity. Moreover, the physical properties being dependent on the atomic arrangement are unique concerning the crystalline and the amorphous state. Moreover, the physical properties, depending on the arrangement of the atoms, are unique with respect to the crystalline and amorphous states. The mechanical properties are significantly different for the two states; for example, an amorphous Pd band with 0.005 cm thick is relatively ductile and stronger and will be plastically deformed under a sufficiently strong bend, whereas a similar crystalline strip of the same composition is brittle and weak and will break with the same bending.

The magnetic and electrical properties of the two states are also different: the amorphous state passes to the crystalline state by heating to a predetermined high temperature with the release of the crystallization heat. Cooling of molten metal to a glass is quite amazing and different from cooling it to a crystalline form: in the first case, the liquid solidifies continuously over a range of temperature without a discontinuous evolution of a heat of fusion, while the crystallization is a thermodynamic first order transition associated with a heat of fusion and a specific temperature.

New interesting compositions were composed mainly of Fe, Ni, Cr, Co and V. Although some compositions, for example, Fe PC Fe Co Co PB Fe BC and Ni PB, have previously been described as being quenched from a melt to an amorphous state, it has been found that some new, excellent and useful compositions can be obtained by adding small amounts (0.1 to 15 atom%, but preferably 0.5 to 6 atom%) of some elements, such as Al, Si, Sn, Sb, Ge, In or Be, to such alloys. These alloys become much better glass formers, that is, an amorphous state is easier to obtain and, moreover, is more thermally stable. It was found that the incorporation of small amounts of some elements of Al, Si, Sn, Ge, In, Sb or Be in amounts from about 0.1 to about 15 atom% to the alloys of the group consisting of Fe, Ni, Co, V and Cr; and Y represents elements from the group consisting of P, B, and C. Excellent glass-forming alloys are obtained. The selected alloys can be relatively more consistent and more easily quenched to an amorphous state than previously thought with known Fe-Ni-Co based alloys. Moreover, these alloys are more stable.

Powders of amorphous metals containing the particles of 0.001 to 0.025 cm can be obtained by spraying a molten alloy into the drops of this size, and then quenching drops in a liquid coolant (water, chilled brine, nitrogen). These alloys may contain small amounts of the other elements contained in the commercial Fe or Ni alloys as the primary source of metals. Such elements, for example, Mo, Ti, Mn, W, Zr, Hf, and Cu, may be added subsequently. The cooling of the molten stream to form an amorphous metallic wire was achieved by spraying the molten stream into stationary water or a cooled solution. The cooling rate experienced by the stream or the jet of molten metal during the quenching depends on the cooling method used for the molten stream and the diameter of the jet. The method of cooling determines the speed at which the heat is removed from the surface of the jet. The diameter determines the surface-to-volume ratio and, therefore, the amount of heat to be removed per unit area to cool the melt in a predetermined amount. For example, it is necessary to meet the requirements by the cooling method, the diameter of the jet and the alloy composition to obtain an amorphous metal wire [23-25, 37] in accordance with the invention include that process described in Patents [23-25].

These amorphous alloys and wire products have valuable physical properties: high compressive strength, elasticity, good corrosion resistance, unique magnetic properties in various selected compositions, etc. In a number of compositions, it is extremely ductile in the amorphous state. Some samples can bend over radii of curvature, less than their thickness, and can be cut with scissors. In addition, using these plastic samples, a tensile strength of up to 350,000 pounds per square inch was obtained under hardening conditions. Thus, thermal treatments often give crystalline materials to obtain high strength, are eliminated by amorphous metal alloys. Amorphous alloys provide a durable material that is resistant to corrosion; selected compositions of these amorphous alloys are relatively stable to the concentrated sulfuric, hydrochloric or nitric acid. For example, the amorphous Feg Ni P B Al is several orders of magnitude less reactive than the stainless steels with the concentrated hydrochloric acid [30]. The preferred novel compositions are characterized by the formula M Y Z [30] wherein M is a metal from the group of iron, nickel, cobalt, chromium and vanadium and mixtures thereof; Y is an element from the group of phosphorus, boron, and carbon, and mixtures thereof; and Z is an element from the group of aluminium, antimony, beryllium, germanium, indium, tin and silicon and mixtures thereof, and wherein the relative proportions in atomic percentages from about to 80, B from about 19 to 22, and C from 1 to 3. These amorphous metals may be suitable for a wide range of applications, e.g. alloys with M iron (e.g. Fe P C Si Al) are of particular interest due to low cost and relatively high strength; Ni Fe P B Al are of significance because of easy formation in combination with high strength and corrosion resistance; alloys with high chromium content (Cr P B Si) have the exceptional hardness and corrosion resistance.

Bulk metallic glasses (BMG) are alloys that can be solidified into a diameter larger than 1 mm without detectable crystallization. The amorphous solid state satisfies the thermodynamic definition of a glass so that heating above a glass transition temperature leads to reaching a metastable super-cooled liquid region before crystallizing. There are many known methods for producing amorphous metals [23-35]. The material properties and the ease of manufacturing amorphous metal specimens depend on the methods used, e.g. the invention [23] relates to a granulating a stream of molten metal, which falls from a launder down into a liquid cooling bath contained in a tank. The metal stream divides into droplets in the liquid cooling bath and the droplets solidify and form the solid granules. The cooling liquid has substantially uniform flow across the tank in a direction perpendicular to the falling metal stream. The flow of cooling liquid has a velocity of less than 0.1 m/sec. The distance from the outlets of the launder to the surface of the liquid cooling bath is kept less than 100 times the diameter of the metal stream measured as the metal stream leaves the launder. From Swedish Patent No. 439783 it is known to granulate FeCr by allowing a stream to fall down into a water-containing bath wherein the stream is split into granules by means of a concentrated water jet arranged immediately below the surface of the water bath. This method yields a rather high amount of small particles. In addition, the risk of explosion is increased due to the possibility of trapping water inside the molten metal droplets. Due to the very turbulent conditions, the number of collisions between the formed granules is high increasing the risk of explosion.

The Iscor Saldanha Steel installed the Granshot™ granulator to accommodate an excess iron production from the Corex plant [28]. The granulation of iron has proven to be the most cost-effective method compared to the traditional alternatives such as the sand bed pooling. The operational cost of the granulator is substantially lower compared to pooling. And the prime metal product showed solid performance as a raw material feed into the steelmaking operations. The benefits were reported due to an increase in stability too. Also the use of a metallic feedstock with a high level of Si and C and without any content of the gangue or unreduced oxides enhanced the steelmaking operation.

## 4. The characteristics of granulated cobalt powder

It is known [13] that a decrease in the cobalt particle size has a dramatic effect on the processing behavior of submicron WC/Co materials. As finer cobalt powders are utilized, the issues of dust (air-borne particulate) generation and environmental stability (i.e. oxidation resistance) are a concern. To address these issues, the cobalt powders were granulated with an organic binder to produce a free-flowing product with a reduced level of dust generation and improved oxidation resistance. The effects of granulated cobalt in the processing of superfine WC powders were examined and compared to nongranulated cobalt powders. Cobalt is commonly used in the manufacture of WC hard metals for a variety of applications such as cutting tools, drills, seals, mining and masonry tool bits. In order to take advantage of these WC powders, finer cobalt powders must be used to ensure the production of homogeneous, high-quality WC/Co materials.

The advantages of finer cobalt powders have been summarized elsewhere [13, 14]. The examined were an Ultra Fine, Sub Micron, Extra Fine HDFP and 400 mesh powder grades, which correspond to cobalt powders with a size of approximately 0.4, 0.8, 1.5 and 4.0 mkm, respectively (Table 1):

Table 1. Characteristics of the cobalt powder grades used in this study

|  | Ultrafine | Submicron | Extra Fine HDFP | −400 mesh |
|---|---|---|---|---|
| FSSS (um) | 0.4* | 0.8 | 1.5 | 4.0 |
| $O_2$ ( %) | 0.9 | 0.6 | 0.2 | 0.4 |
| $D_{50}$ (um) | 4 | 5 | 7 | 15 |
| AD (g/cc) | 0.6 | 0.8 | 1.1 | 2.2 |
| S.A. ($m^2$/g) | 4.5 | 2 | 1 | < 1 |

Many different methods are commercially available to granulate the fine powders [16-18]. During granulation, a force acts on the powder to bond it together into larger granules. The forces bonding the powder together can be classified into three categories: 1) capillary forces generated when a liquid is used to granulate a powder together; the strength of these granules is generally low; 2) adhesive bonding; a solidified binder phase adheres to the surface of the powder bonding particles together into larger granules; 3) the third force for granulation is solid bridging. The binder phase is usually an organic material (paraffin wax, PVA, etc.). The strength of granules is dependent upon the amount and type of the binder phase present. Generally, the strength of adhesively-bonded granules is higher than the ones formed by capillary forces.

After the discovery of the first metallic glass (1960) [1], the amorphous metals were made in a variety of compositions, mostly fabricated as thin ribbons less than a millimeter in thickness because fast cooling rates (up to $10^6$ K/s) have been required to retain the metastable amorphous phase. Recently, a new class of the amorphous metals developed has required the cooling rates of only 1 K/s [6]; such alloys, as Zr41.2Ti13.8Cu12.5Ni10Be22.5 (at%), can thus be processed in bulk form. The amorphous metals generated much interest [26], both in basic research and structural applications, because of their near-theoretical strength to stiffness ratios and extremely low damping characteristics [7]. What is more, a number of amorphous metals revealed high corrosion resistance [8-12], which was explained by their structural homogeneity. It was also examined [26], whether the amorphous condition itself confers improved corrosion resistance.

An extraction step to the final refining in the pyrometallurgical smelting process requires melts casted and/or comminuted for further taking up into solution or leaching [22] using mostly the hydrometallurgy. A general practice with a friable material having the properties associated with the higher sulfur, carbon and silicon levels is to granulate and in many cases to mill them, to develop a surface area for the optimal leaching. But in case of the metallic melts, a milling can be unfeasible thanks to the ductility of the granules. Water solidification is applied for producing the fine particles through a process of the granulation or Atomization. However, the Atomization can produce particles of about 40 mkm size without a milling circuit. Hydrometallurgy has been used for centuries in the precious metal refining but it is now widely used for the metals like Co, Cu, Ni, Ta, Nb, etc. Maximum powder sizes (250 mkm) and median sizes (40-75 mkm) are normally preferred. Atomization is widely used for processing aluminium, zinc and copper alloys at rates of 0.5-3 t/h.

As was mentioned in [22], the main concept of liquid spraying (Atomization) was known to Agricola who granulated a molten metal already in the 16th century. In the period of the development of capitalism in the XIX century, the Atomization of tin and solder was introduced for the manufacture of pastes for plumbing. And then, hundred years ago, Prof. Hall in the USA patented spray nozzles for aluminium, applied in explosives (1915). He used compressed air or the easily accessible steam for spraying. The air is still mainly used for the processing of aluminium, zinc and copper alloys at a rate of 0.5-3 tons per hour. Works on the processing of copper alloys and cast iron were carried out using air and water jets in the 1930-1940 years. Dr. Jones was a pioneer in the treatment of Fe, Ni and Co alloys (England, 1940-1950). After the launch of commercial applications, a large scale (10-30 ton batch) water Atomization of the carbon steel powders was carried out. Now there are factories with a capacity of up to 50 t/h (world production is about 1 Mt/y) and several factories more than 200 thousand tons per year. This large-scale, low-cost melt processing technology is now commonly used in the metallurgical industry, especially for the high-temperature melts. Our methods based on the controlled jet and film flow disintegration are more complex and allow producing the small particles in a strictly defined size (in a wide range of the sizes available) but it is available only slightly over $10^3$ Celsius degrees. Over this temperature, we did not solve a problem with the durable materials for our granulation machines, for the moment.

Atomization allows the conversion of the liquid melt to powder in one fairly straightforward operation. Apart from the advantage of having one less step in the process compared to granulation and milling and four less compared to casting, crushing and milling, there is also the advantage of dispensing with all attrition components in the milling process. Therefore, it is attractive to consider atomization to replace milling and grinding when extraction of Cu, Ni, and Co from sulfide ores and to use similar technological ways to extract metals from the landfills Cu, Ni, Co, where slag is processed through smelting. Alloys obtained during this reprocessing are low in sulfur and hence ductile, and are not suitable for processing by crushing and/or milling. Obviously, while atomization with high-pressure water is generically related to water granulation, the much higher water pressure (20-200 bar compared to the 2-6 bar for granulation) significantly affects the design of the equipment. It's expensive to pump all the water needed to extract heat from the melt without boiling (usually 5 or 6 times the mass of the metal). Using special designs, to spray the melt, you can use a high-pressure water stream only 2-3 times, and for cooling - additional water of low pressure.

Granules can be handled by front loaders, magnets or a conveyor to introduce them into the metallurgical process, preferably continuously or periodically. The granular product has a high bulk density, usually 3500-4500 kg/m$^3$, for penetration into the slag layer during the addition, while the shape and size provide for rapid melting and dissolution in a hot metal melt. The generation of fine particles, defined as a material less than 4 mm, formed during the processing, packaging, and loading of FeCr producers, as well as during transportation has been studied in [21]. Such should also take into account the formation of fines during re-packaging in the intermediate storage and, finally, fines generated by the end user at the expense of internal transport, processing, storage and introduction into the metallurgical process. The comparative data to describe the difference in the fines, from handling and transporting the purchased HC FeCr product by 10-50 mm in comparison with the product of granular HC FeCr was made. The size of the granular HC FeCr was the direct output of the GRANSHOT process [21]. These granules were to almost 99% over 4 mm size, no dust generation. The weight and size, as well as the homogenous and clean analysis of the granulated material, are ideal from the logistical and metallurgical point of view. Some large HC FeCr granules were during tumbling broken into two pieces but with no major fines generation. Tumbling of crushed HC FeCr generates fines and losses at some 10 times higher levels compared to granulated HC FeCr. The combined figure of material losses and fines (<4mm) was reported a total of 7% for crushed material as compared to 0.7% for granulated one. The tests in did not represent an actual generation of fines in the industry during production, transport or at end-users, but served as a good indicator of a significant difference between the crushed and granulated HC FeCr when it comes to dust and fines generation. The results [21] comply well with UHT experience of the increased production and use of granulated ferroalloy.

The need for increasingly better steel grade products and higher quality levels requests the raw material, such as granulated ferroalloys which has a homogenous composition, a minimum of pollutants or oxides and rapid melting properties. As many ferroalloy grades, especially low carbon grades, LC FeNi, LC FeCr, are added at the very final stages of the steelmaking process, then requirements on cleanliness become even more important, because little time for removal of impurities is available. The granule size and shape also fit well to rapid and complete dissolution in the melt.

## 5. Our field of study by the fine precise granulation of the middle-temperature metals

In a contrast to the above-considered works by amorphous materials and granulation of the metals and alloys, we deal with a granulation of the middle-temperature metals (Cu, Al, Pb, Mg, etc.), with the substantial request of the spherical form of particles and strictly uniform size (mainly below 1 mm). The methods and devices for production the spherical granules (particles) of metals of a given size, with a high cooling rate by solidification are presented. The cooling rate of the drops during their solidification has been achieved up to $10^4$ Celsius degrees per second in our experiments! Such fine structure metals are amorphous or close to amorphous metals. The idea of their creation appeared on the assessment of the iron's hardness by Academician Ya.I. Frenkel nearly hundred years ago (1925) [39]: theoretically the iron strength differs from the real one up to thousand times in some of its properties. The conventional production methods of the metallurgy lead to a deterioration of all properties of the metals due to slow cooling of a metal from its liquid to a solid state and a forming of the big crystals due to this instead of the amorphous structure.

The granular technology in the material science requires amorphous particles of nearly the same size (the uniform properties) produced from liquid drops, cooled down with a high cooling rate so that to avoid development of the big crystals, which can decrease the quality of a metal dramatically. The uniform particles fit the best to further production of the new material because from different particles the material produced has nonuniform special properties. This may decrease a lot the advantages of the amorphous final product. The granular technology is a contrast to the powder metallurgy, where cooling of particles of different size is usually going with a comparably low cooling rate being. Therefore granules of a small size are normally required, except some special applications other than the ones of the material science. Big spherical granules cannot be produced this way because the big liquid drops have non-spherical form and worse conditions for the rapid cooling during solidification. The capillary forces fall rapidly down by growing of a diameter of particles over 3 mm or so.

Our results in a creation of the highly efficient processes and the perspective technological devices have been based on the earlier developed theory of the parametric control of film flows, as well as on the discovered new phenomena of the controlled disintegration of film flow into the drops with a further rapid solidification of the drops [40-43]. Some of these basics are presented in the paper. The scientific novelty of this work consists in a development of the principles and the optimal regimes in a production of the amorphous granules for the creation of the super hard and the functionally unique composite metals. Different possible technical and technological applications are presented and discussed here for the new methods developed by us.

The new theory of the parametrically controlled jet and film flows has been developed by our team at the Insitute of Electrodynamics of Ukrainian Academy of Sciences [40-54]. The technological processes based on the discovered phenomena and the invented methods and devices for the materials' granulation [44-48] were patented and implemented into metallurgical and space industry. Here we present a description of the film flow granulation machine for Zink, Aluminium, Magnum (Zn, Al, Mg) and their alloys, which has no analogs in the world practice. In a development of this granulation machine the following three important problems have been solved: parametric control of a disintegration of the liquid metal film flow (a dispersing knot), stabilization of the phase transition boundaries (protection of the channel walls against destruction and, at the same time, protection of a liquid metal in a channel against pollution with particles from the channels' walls), selection of an optimum cooling regime for the granules with achievement of a highest cooling rate for the drops during their solidification into granules. The first one was realized in a film granulation machine, while the other two have got an only theoretical solution for the moment.

## 6. The parametrically controlled disintegration of liquid metal film flow into the drops

The most important results obtained, which have been created a base for the brand new perspective granulation machines and have theoretical interest for controlling the film flows, are as follows:
1. The theory and engineering methods of the parametric excitation and suppression of oscillations in the film flow under an action of the electromagnetic fields and vibrations.
2. The theoretically predicted and experimentally tested three new phenomena of the disintegration of the liquid film flow into the drops using the alternating electromagnetic fields and vibration:
    - the resonant electromagnetic decay of a film flow into drops of strickly given size,

- the vibration soliton-like film flow disintegration into the drops in a narrow range of the drops' sizes (nearly 50% of the average size)
- and the shock-wave regime for the film flow's disintegration into the microsize drops with a narrow size distribution.
3. The experimental facilities for a rapid cooling of the drops during their solidification, with up to $10^4$ Celsius degrees per second, which allow producing the amorphous nanostructured granules for the new materials' production with the unique properties.

These results are new in a field of the metals' granulation, which may serve for a production of the new amorphous materials, as well as for development of some materials with the unique complex structure by application of the electromagnetic control of the flow inside the liquid drops (maybe with some adding) before their solidification. It is also interesting for other purposes, both in a theory and in practical applications. They may be also useful for the scientists and engineers working in a development of a theory of the controlled film flow, in constructing and implementation of the granulation machines and dispergators for the industry, etc.

The operation principle of the granulation machines for electromagnetic batching of metal alloys [49-51] with a melting temperature approximately no more than $10^3$ Celsius degree is based on a controlled disintegration of free the jet flow on the first mode of its Eigen oscillations. It is producing the metallic granules of the size from 2 mm and over from the metal melts and alloys. Applying the crossed electric and magnetic fields at the exit from the channel (nozzle) delivering a melt leads to the resonant regime. A frequency of the alternating ponderomotive electromagnetic force is chosen equal to a frequency of the jet's Eigen oscillations. The magnetic field is created by a permanent magnet as the simplest way but can be arranged any other method. A current in a melt at the nozzle has the industrial frequency of 50 Hz. A number of different constructions have been developed and successfully implemented in industry and research facilities. The crystallizer was done depending on the granulating melt and requirements as a concern to a size of particles and cooling rate. Coolants were air, water, water-cooled copper plate, etc. As a jet's disperser it was also applied for creation of the liquid metal protection curtain in an experimental tokamak.

The electromagnetic batcher implements the controlling capillary disintegration (fragmentation) of a cylindrical jet of liquid metal in the range of 100...400 Hz (electromagnetic force has 100 Hz frequency, created by applying the electric current of 50 Hz). The diameter of the metal spherical particles (granules) is determined by the resonant frequency of a jet at the first harmonics, which is approximately $\omega=0.23u/d$ for the low viscous melts [40, 49, 51]. Here $u$ is the flow velocity and $d$ is the nozzle's diameter. A diameter of the spherical particles, which are produced due to a jet's decay, is roughly estimated as $d_p=1.88d$. Because the capillary forces are growing with a decrease of the nozzle's diameter (inversely proportional), the controlled disintegration of a jet is available up to the nozzle's diameter of about 1 mm. Afterward, the Coanda effect makes the jet chaotically directed flow from the nozzle. It is breaking the jet's strictly vertical flow. In addition, even an accidental slight clogging of the melt with some impurities can also destroy the jet's flow regime and cause the granulator to stop. Therefore, the jet's type granulators produce particles of a size over about 2 mm.

Thus, the jet granulators, despite their advantage in getting nearly ideal spherical granules of the same given size, revealed serious impediment to the size limitation about 2 mm. The granular technology requires particles below this limit, sometimes substantially below 1 mm. Therefore the theory of parametrically controlled disintegration of the liquid metal film flow has been developed [40, 42], and the methods and devices for film granulation were invented [45-49]. The last ones produce granules in the range of 0.5-1.5 from the average size, but they practically do not have any limitations on a size of the producing granules up to the micrometer size. We have developed the theories of liquid metal film flows and their parametric control for disintegration into particles of given size [40-43]. Diverse phenomena of the parametric excitation and dumping of oscillations, their application especially for the case of rapidly spreading films on the surfaces have been studied from the beginning of 1980[th]. The results of theoretical and experimental investigations allowed designing and successful testing of the new prospective granulators of metals, as well as of some other devices, for the new material science and engineering.

## 7. Forming the thin film flow as a result of jet's shock on a plate

Forming of the thin film flow is the following fascinating phenomenon. Vertical jet transforms into a horizontal thin film flow. The velocity of a jet is substantial but a disk is immovable. Let simplify the task for the flat jet, which actually gives the result close to the round jet. For the shock of a vertical jet on a horizontal disk, the momentum conservation law must be considered. The mass of an amount of jet's liquid, which is abruptly decelerated from the high velocity of a jet to a zero velocity of a disk, during small shock time $\Delta t$ is $\rho c \Delta t S$, where $\rho$ is density of a liquid, c is the speed of sound in this liquid medium (e.g. for water it is about 1500 m/s), S is the area of a cross-section of a jet. Due to the symmetry of a flow, only a half of the jet and film flow is considered. This mass multiplied by a jet's velocity yields a momentum of the decelerated part of a jet due to a shock with the disk. The momentum of the decelerated part of a jet is equal to the impulse of force created by jet on the disk due to a shock: $PS\Delta t$. Then P is computed as the pressure due to collision of disk and jet.

The momentum conservation equation for the shock of the moving jet with the immovable disk is $P \cdot b \cdot \Delta t = \rho \cdot c \cdot \Delta t \cdot b \cdot u_{00}$, where $S = b \cdot 1$ (width of a jet multiplied by a unit distance in a direction perpendicular to the plane of the figure, y). Then $P = \rho \cdot c \cdot u_{00}$, so that it is much higher than a normal action of the jet on a plate by a smooth flow when it is equal to a kinetic energy of the jet, which is $P = 0.5 \rho \cdot u_{00}^2$. For example for water jet with velocity 1 m/s the pressure in a shock is about $1.5*10^6$ N/m$^2$, which is 3000 times higher than the normal pressure from the same jet flow caused on a plate (500 N/m$^2$), without shock, by a smooth flow. Similar phenomena are also encountered in other situations, for example, waves in a storm, beating against a barrage on the seashore and jumping much higher than the waves in the ocean.

Using the above, we can compute velocity in a film flow created by a jet after its shock with a disk. The momentum conservation for the film flow is: $\rho \cdot c \cdot u_{00} = 0.5 \rho \cdot u_0^2$, where from follows

$$u_0 = \sqrt{2c \cdot u_{00}}, \quad a = b \cdot u_{00} / u_0, \text{ or } a = b\sqrt{0.5 u_{00} / c}. \qquad (1)$$

In the above conditions, it is computed: $u_0 \approx 55 \ m/s$, $a \approx 0.018 \cdot b$, where from yields $a \approx 0.09$ mm for $b=5$ mm (a jet of 1 sm width, with velocity 1 m/s creates less than 0.1 mm thick film flow with a speed 55 m/s!). The viscous dissipation of the flow energy is not taken into account, therefore these values are overestimated but still exciting.

In the film flows, the ratio of various forces significantly depends on the film thickness, physical properties of liquid metal, type and intensity of the external influences that define a big variety of the flow modes. A noticeable role is played by the capillary, viscous, electromagnetic forces [40], etc. As mentioned above, the moving liquid film can be disintegrated into the drops of given size depending on a frequency of the parametric excitation (e.g. alternating electromagnetic field), with the comparably low energy consumption. In the other case, the system of solitons throws off drops like a series of the unit jets do in a phase of vibration modulation. The third case is available with a shock wave on a jet, at the high vibration Euler numbers (10-100 depending on the parameters).

## 8. The processes of the controlled jet and film flow and their fragmentation into drops

Controlled fragmentation of the jet and film flows and the drops' formation due to this is of interest for metallurgy, chemical technology, new material science, etc. What is more, these phenomena must be considered altogether because the real processes are complex, e.g. solidification of the drops, flow of a coolant through a granular layer, form of the solidified granules and their distribution by sizes in the range stated. Nowadays the problem is stated not only to develop the models and to simulate the processes but also to control them according to the goal stated. Parametric control in the heat and fluid flow is important in practice for the following reasons:

- Parametric jet and film flows' disintegration: dispersing and granulation of the metals (producing the granules), spray-coating, melt spraying.
- Stabilization of the unstable processes with the parametric control: the jet/drop and film screens for protection of the diaphragm in the experimental tokamak, the stability of the plasma and combustion, reduction of the hydrodynamic friction, and the others.
- Parametric excitation of the oscillations for intensification of the heat and mass transfer processes, as well as for improving the mixing and the quality of a crystallizing metal.

The parametric control allows, in some cases, running the processes, which are unstable in an absence of the corresponding stabilizing action (suppression of perturbations). Examples of the parametric disintegration of the jet and film flow using the vibration is shown in Fig. 1, where the multiple series of the drops are produced after the parametric disintegration of the jets on a vibrating horizontal plate:

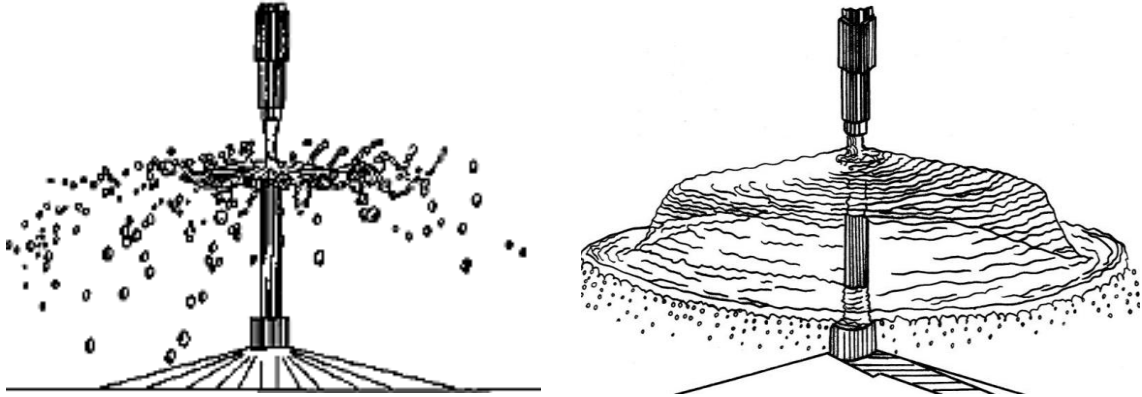

Fig. 1 The disintegration of the jet and film into the drops under vibrations of a disk

By the low vibration Euler numbers, the film flow decay is ineffectively controlled (Fig. 1 to the right). The drops are produced from the edges of the vibrating film flow. Under such conditions, only some regularization of the process is available, e.g. twice narrowing the drops' sizes distribution.

According to the above, a radial film flow is formed as a result of a vertical jet 1 transforming into a film flow on a horizontal plate as presented schematically in Fig. 2. A film flow is considered under control by an alternating electromagnetic field, which is applied in the vertical direction. The film flow is spreading over the horizontal disk. The parametric wave control in a film flow is organized by an action of the progressive electromagnetic wave in the following form:

$$h = h_m(z,r) \exp i(kr + m\varphi - \omega t), \qquad (2)$$

which is created by the inductor 2 installed both over the film flow and below the disk plane. The wavenumbers $k, m$ by coordinates $r, \varphi$ and a frequency $\omega$ of the electromagnetic field can be regulated, $h$ is a vertical component of a strength of the magnetic field, $h_m$ is its amplitude. Also, the vertical harmonic vibration of the horizontal plate is considered as the other parametric control in a film flow:

$$d^2z/dt^2 = g_v \cos \Omega t, \qquad (3)$$

where $g_v$ is the acceleration due to vibration, $\Omega$ is the frequency of vibration.

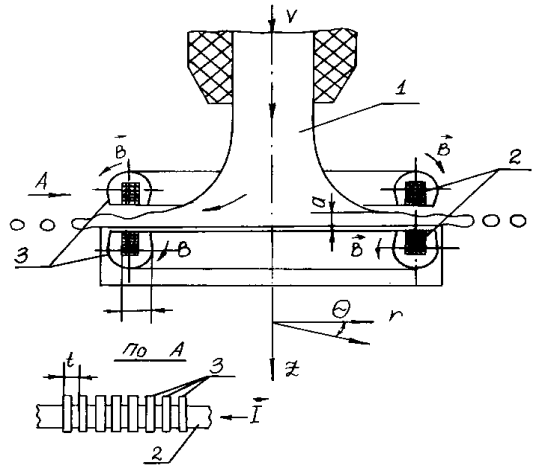

Fig. 2 Electromagnetic inductor for the disintegration of the film flow into the drops

The parametrically controlled disintegration of the film flows into the drops of given size was studied on the mathematical models developed using the computer simulations, as well as the experimental investigation of some regimes was done. The parametric control was considered by the electromagnetic field and vibrations and by their common application too. This is one of the most modern actual problems in the theory of heat and fluid flow control. And its applications to the modern industry (chemical technology, special metallurgy, perspective material science and many others) are highly promising. The other two applications of the parametric control are somehow opposite to the controlled disintegration of the jet and film flows into the drops. They have a concern to a suppression of the surfaces' oscillations of the film flows with a goal of their stabilization in the unstable regimes if any. Also an excitation of the given parameters' oscillations for an intensification of the heat- and mass- exchange in different processes was intensively paid attention since the last decades while some of the parametric control processes were studied and applied over 100 years ago and even more. For example, electromagnetic action on a quality of the solidifying melt was first considered over hundred years ago and studied intensively during the last few decades.

The film granulation machines invented and tested by us [44-54] are unique in the world up to the present time. They are distinguished by simplicity, the developed specific surfaces of a melt and an intensive cooling of the granules in the liquid nitrogen film flows. These granulation machines allow producing particles of the strictly given sizes in the different regimes depending on a frequency of the parametric action, a flow velocity at the nozzle and a diameter of the nozzle. There is a possibility to produce the amorphous identical particles due to a high cooling rate, in a narrow range of particles' distribution by the sizes. In contrast to the earlier developed the jet granulation machines, the film granulators allow obtaining granules practically of any desired size, with a narrow distribution of particles by sizes (deviation by size just no more than plus-minus 50% from the average size).

Actually, the film flow devices considerably surpass the traditional ones with a working liquid body being ecologically clean too. What is more, much high-efficient withdrawal, heat- and mass-transport, and the other processes and installations can be constructed on a basis of the film flows, especially in a case of the parametric film flow control.

## 9. Film flow dispergators and granulators with electromagnetic and vibration control

General view and schematic construction of the film granulator with vibration control [54] are demonstrated in Fig. 3. The main elements of the construction are: 1- vertical channel for the transporting of the liquid metal from furnace to the vibration disk, 2- nozzle, 3- vibrating disk connected to the vibration machine under it, 4- bar connected to the vibrator, 5- membrane of the vibrator (was chosen to resonance vibration, could be changeable for getting the desired resonance frequencies), 6- vibrator, 12- internal box inside the working chamber of the granulator filled with a liquid nitrogen, 13- cooling system with liquid nitrogen.

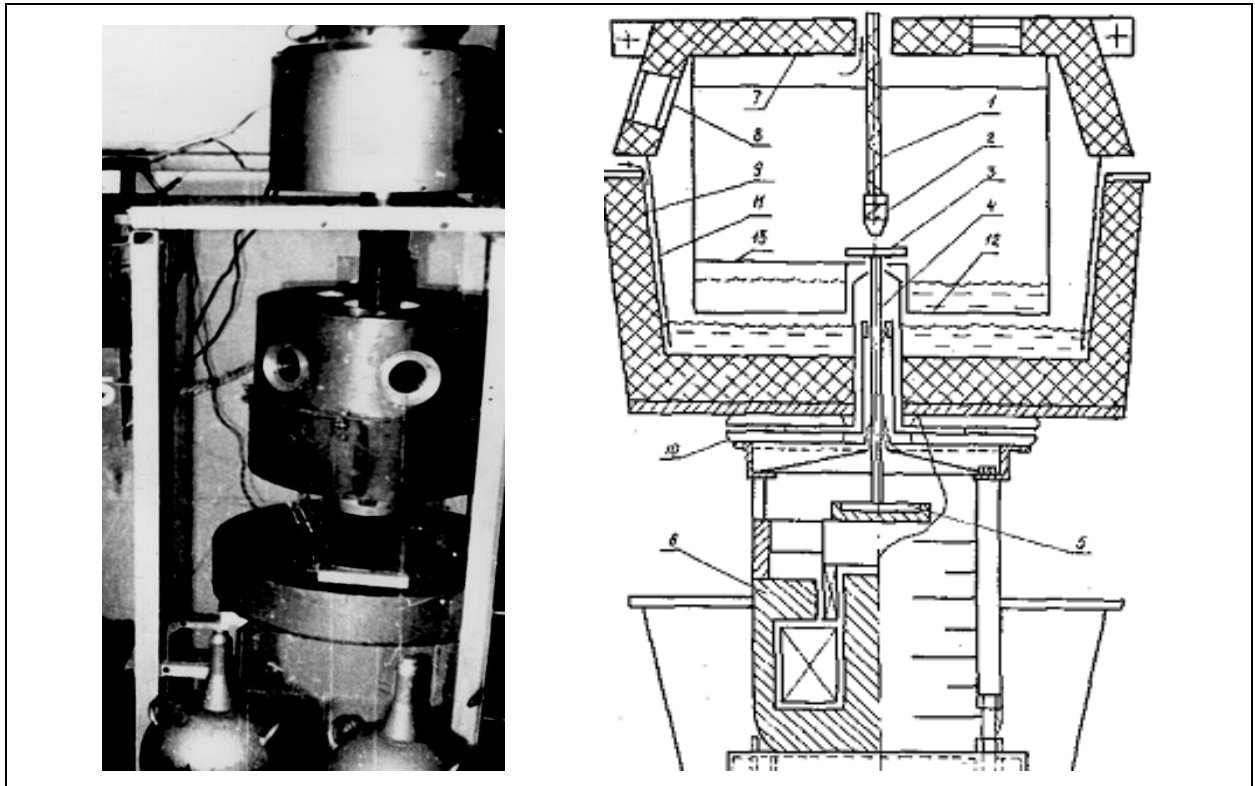

Fig. 3 Facility with vibration type controlling device for the disintegration of the film flow, with a nitrogen atmosphere and liquid nitrogen coolant for the drops

Experimental study of the controlled film flow disintegration into the drops of given size has been done using the invented and constructed facilities. Two of them have had electromagnetic control devices, while another two had the vibration type controlling devices (with deep vacuuming of the chamber and with controlled nitrogen atmosphere in the chamber of the facility). The uniqueness of the granulation machine includes the special vibration type control unit, which can work using the standard low-power vibrator modified by application of the resonance membranes for different vibration frequencies. Granulation machine was built on the principle of the parametrically controlled disintegration of film flow in the optimal modes of parametric oscillations, which were revealed. Nitrogen atmosphere in a chamber and liquid nitrogen cooling system for the granules in the form of a series of liquid nitrogen film flows below the liquid metal film flow have been invented to avoid the heat transfer crisis due to nitrogen vaporization around the drop (granule), so that to keep the highest intensity of cooling during all process till the complete solidification of the granules.

After the disintegration of a film flow into the drops, in a narrow range of the given size, the drops are moving through a series of the film flows of liquid nitrogen (the bell-type film flows), to avoid the heat transfer crisis. Under such conditions, the cooling rate for the drops and particles achieved 10 thousand Celsius degrees per second (a world record!) so that the amorphous structure of the metal was obtained. The cooling rate of the million Celsius degrees per second was achieved for the flakes in the USA (the Al drops were shot with a high speed on the water-cooled Cu plate). But the flakes are not as good as the uniform granules of a nearly spherical form are for a production of the amorphous materials. Erwin Mayer invented a novel method, which was later on called "Hyperquenching" [55], and then developed it during about 25 years. This was a breakthrough in solving the problem, which the Royal Swedish Academy of Sciences emphasized as the key role of the hyperquenching method and Erwin Mayer in the scientific background for the 2017 chemistry Nobel prize (E. Mayer has often interacted with J. Dubochet starting from the first work [55]). This method allowed achieving the cooling rates of up to 10 million $^0$K/s, sufficient to beat a formation of the ice crystal for pure water, as well as for all aqueous solutions.

We discovered in detail the three new phenomena of the resonant, soliton-like and shock-wave disintegration of a film flow experimentally and theoretically. Then we have implemented the above results in a development of the unique prospective facilities, the so-called film flow granulators, and dispersers. The vibration type control of a disintegration of the liquid metal film applied in the facility shown in Fig. 3. The new phenomena of the soliton-like disintegration of a film flow and the shock-wave disintegration of a film flow have been implemented in this type granulation machine. The first one can be observed in Figure 4, where the drops are levitating over the vibrating plate. It looks like a chaotic process but it is not chaotic indeed. A process is regularly controlled, and the drops are produced from a series of the solitons, which all are nearly uniform and they cover all the plate, which is vibrating in a vertical direction. In this regime, there was also observed cavitations on the vibrating plate, but it was not studied by us for the moment concerning to its influence on the drops' formation.

The shock wave regime was obtained with nearly ten times higher vibration Euler number and looked like conical shock-wave on the vertical jet, which produces fine particles (of mkm size) from the jet. The film flow does not exist in this case. This regime may be used for the production of ultra-small particles or for the spraying of materials. Vibration type devices are applicable to any appropriate melts except the highly viscous ones. The results in Figs 4-6 were obtained using water as modeling liquid because illustration of the controlled disintegration of the liquid metal film flow is not available. Thus, by $r_0 = 3,75 \cdot 10^{-3}$ m, $u_0 = 2,6$ m/s, $T_{water} = 8^0 C$, $T_{air} = 14^0 C$ yields $\Omega_* = 440, 880, 1760$ Hz. By $\Omega_* = 440$ Hz spraying is weak, by $\Omega_* = 880$ Hz - stronger, and by $\Omega_* = 1760$ Hz spraying is going directly from the surface of a disk but there is also some short film flow on a surface of disk (Fig. 4):

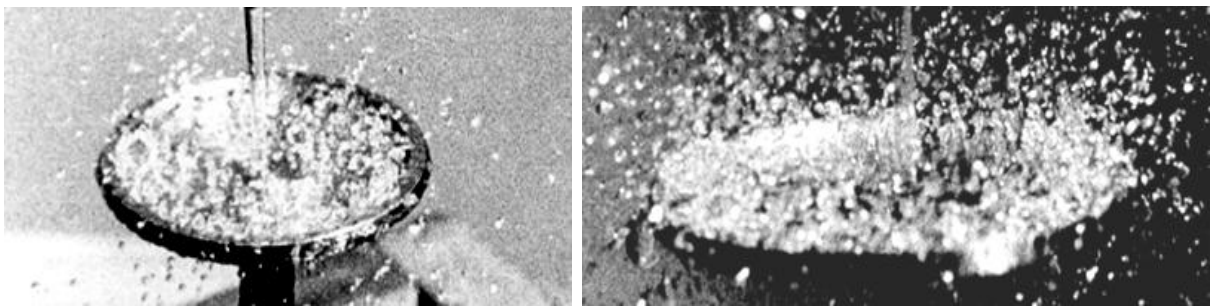

Fig. 4 Start of the vibrational spraying of film flow: $r_0 = 3,75 \cdot 10^{-3}$ m, $u_0 = 2,6$ m/s, $\Omega_* = 1880$ Hz, $We = 342$, $Fr = 13,6$, $\text{Re} = 7,44 \cdot 10^3$ (to the left) and partial vibro-spraying: $\Omega_* = 1750$ Hz, $g_V = 160 g$ (to the right)

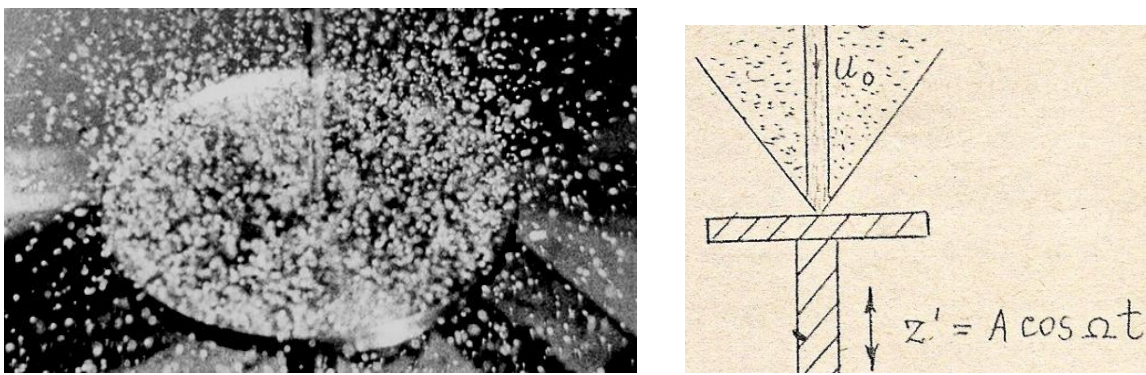

Fig. 5 Vibration controlled soliton-like regime (to the left) and the shock-wave film flow disintegration regime (to the right): $r_0 = 0,8 \cdot 10^{-3}$ m, $u_0 = 0,2$ m/s, $\Omega_* = 450$ Гц, $g_V = 250 g$

By $u_0 = 2,8$ m/s results: $\Omega_* = 490$, 805, 1610, 3100 Hz, and $\Omega_* = 1610$ Hz yields nearly total spraying, while the other regimes give just spray at the edge of film. The regimes of soft loosing of instability show a gradual increase of oscillations' amplitude of a disk, which leads to a monotonous growing of amplitude of the film flow perturbations. At $Eu_g \gg 1$, monotonous increase of vibration acceleration $g_V$ leads to an explosive collapse of a film, its spraying into the smallest droplets invisible to eyes [42, 53]. The conic shock wave is passing to a jet head from a film: the jet turns out with the sharp nose (forming lines of a shock wave) touching a disk surface (see Fig. 5 to the right). The sprayed drops are in the cone formed round a jet by a shock wave.

The mode of high vibration accelerations creates a system of the non-linear waves destroying a film on the drops on a film surface. Thus, the vibration accelerations providing $Eu_g \sim 1$ correspond to a case of soliton-like solutions. At $Eu_g \gg 1$ there is a shock wave passing to a vertical jet, and the film disappears, there is a rigid mode of stability loss (explosive transition of a film to small drops, spraying from a disk surface in a jet contact place). Emergence of solitons at $Eu_g \sim 1$ leads to that dispersion of waves in radially spreading film disappears and separate solitons under the influence of vibrations work as separate vertical jets, from which crests the drops break. For checks of uniformity of disperse structure on the above device with the nitric atmosphere the Gallium granules were produced (see Fig. 6). The particle size distribution of the produced granules cooled in liquid nitrogen (15 sets of granules by 10 pieces were three times considered), are given in the Table, where are: $n$- quantity of granules, $d$- diameter of granules (in mm). The equivalent diameter of granules is

$$d_{eqv} = \sum_{i=1}^{12} n_i d_i / \sum_{i=1}^{12} n_i \approx 0,88 \cdot$$

| $n$ | 12 | 24 | 25 | 19 | 15 | 15 | 9 | 14 | 8 | 3 | 3 | 3 |
|---|---|---|---|---|---|---|---|---|---|---|---|---|
| d | 0,5 | 0,6 | 0,7 | 0,8 | 0,9 | 1,0 | 1,1 | 1,2 | 1,3 | 1,4 | 1,5 | 1,6 |

The dispersive structure is quite uniform, especially with account of the considerable errors of experimental data connected with a non-stationary process (drift of a resonance of the vibrator at dependence on a melt flow rate). Exercising control of drift of a resonance and its automatic deduction, it is possible to narrow the dispersive structure of granules. The developed new principles of a granulation allowed constructing the film MHD- and electrodynamic granulators differing significantly in bigger productivity (1-2 decimal orders) comparing to the known jet granulation machines [49-51]. They don't have analogs in the world, considerable wider range of the produced mono-granules, and simplicity of a design and low-energy consumption. One of the semi-industrial granulators, FGA-1 (the film granulator for aluminium alloys), was successfully tested in vitro and at the enterprise [53, 54]. Works in this direction are perspective for many industries and technologies [40-42, 54, 56, 57], however since 1991 they were dramatically decreased due to absence of financing.

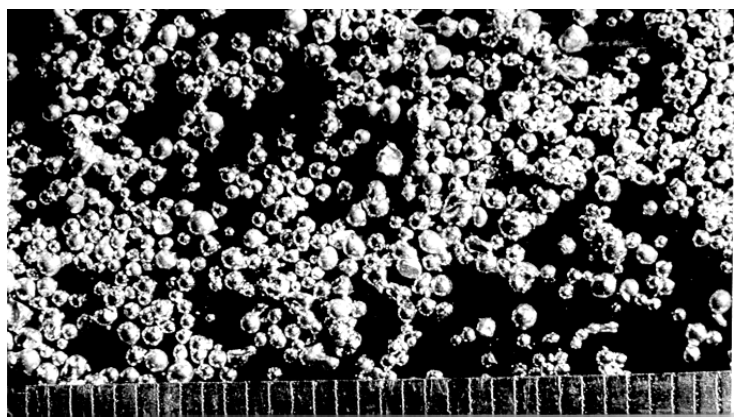

Fig. 6 Gallium granules produced on vibro-granulator FGA-1 (cooling in liquid nitrogen)

In contrast to the free film flow decay, the electromagnetically controlled film flow disintegration [42, 52] was obtained in a modeling of the film flows with the liquid metal Gallium (Fig. 7):

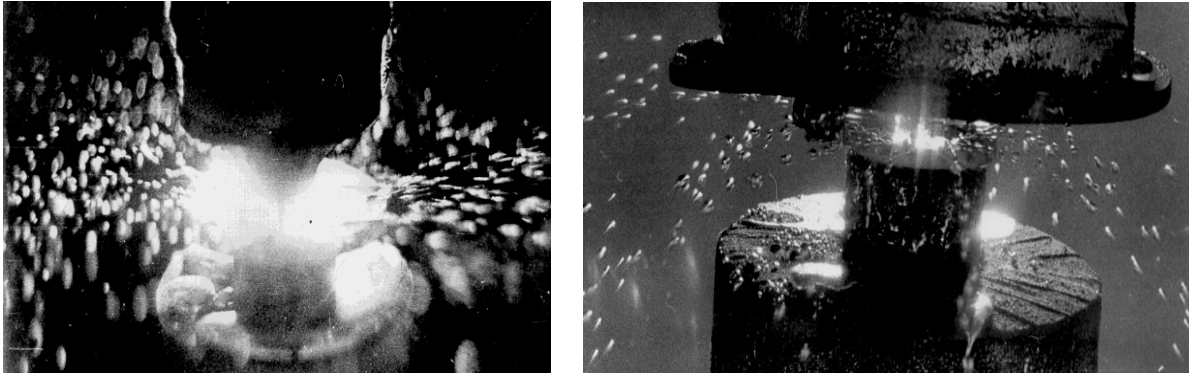

Fig. 7 The free decay of film flow (to the left) and electromagnetically controlled resonant disintegration of the film flow (to the right)

Fig. 7 clearly demonstrates that the drops and granules produced are nearly uniform by their sizes. The drops of controlled size can be produced using the developed by us granulation machines for different metals. The facility was proven on a few metals and alloys with a solidification temperature below $10^3$ Celsius degrees. Granulation was prepared for special metallurgy and for production of the new materials with the unique properties. Distribution of the drops by sizes was achieved narrow, with a deviation of about 50% from the given size as shown in the Figs 8-10.

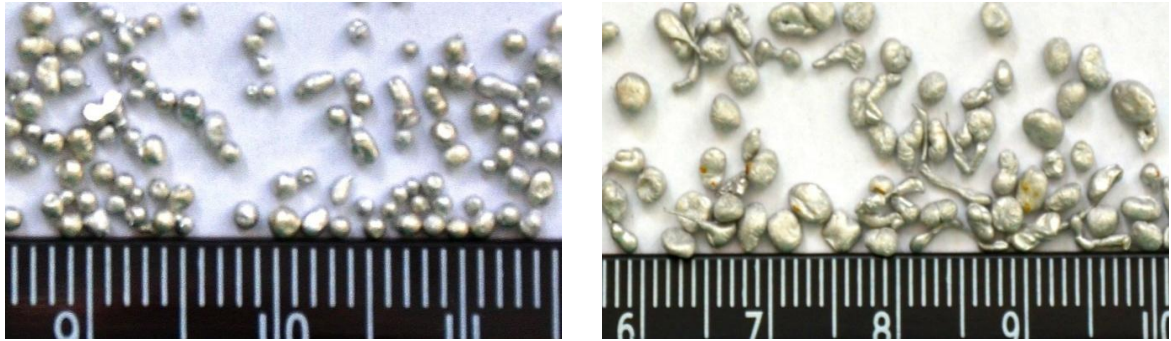

Fig. 8 Particles of Al produced in soliton-like regime (to the left, without separation by size) and on centrifugal granulator (to the right, particles separated in the given range of sizes)

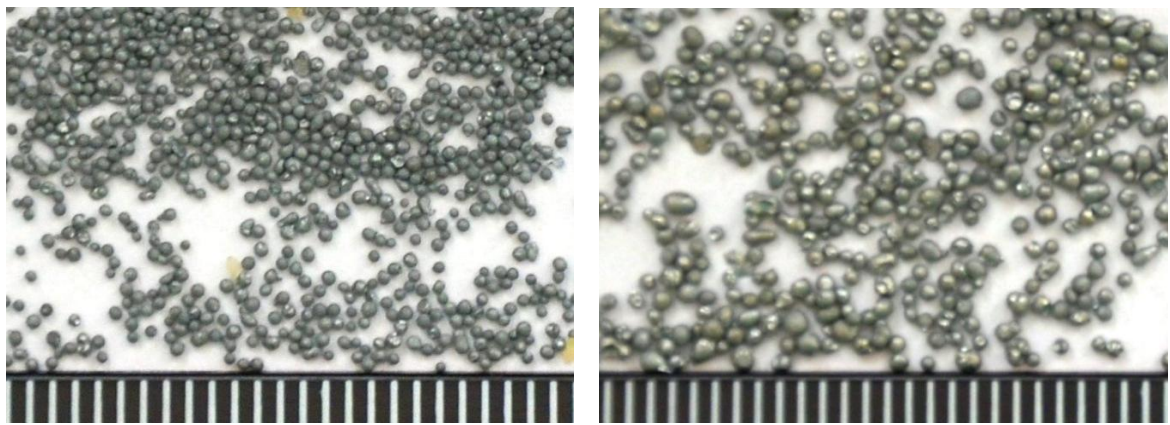

Fig. 9 Particles Zn produced in soliton-like regime on vibration type granulator

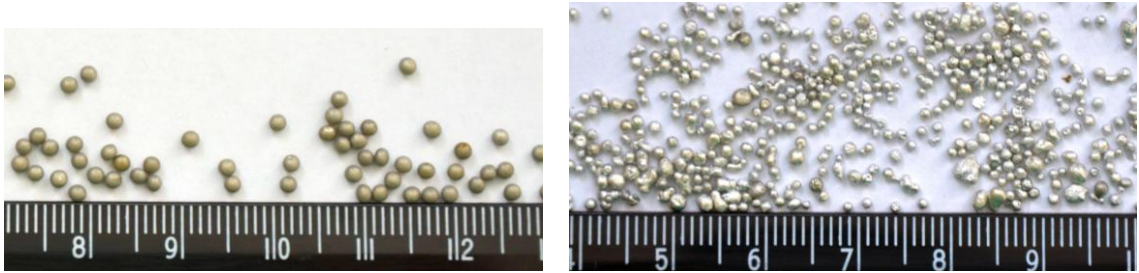

Fig. 10 Aluminium granules produced on the jet granulator (to the left) compared to the granules produced on the film granulator (to the right)

For comparison, a distribution of the drops by sizes in the process of free film decay was broader than the given value 10 times and even more. The vibration granulation machine worked using the standard 10 kW vibrator but with special membranes in the resonance regimes. These membranes allowed obtaining the required vibration frequency and vibration acceleration up to 2000 m/s$^2$. The electromagnetically controlled disintegration of the film flows shown in Fig. 3 was done in a vacuum chamber with the nitrogen atmosphere. The weak vacuum in a chamber (1 mbar for example) did not allow working with some melts due to a very strong thin oxide film on a surface of the liquid metal film flow, which makes the process totally chaotic as seen in Fig. 11:

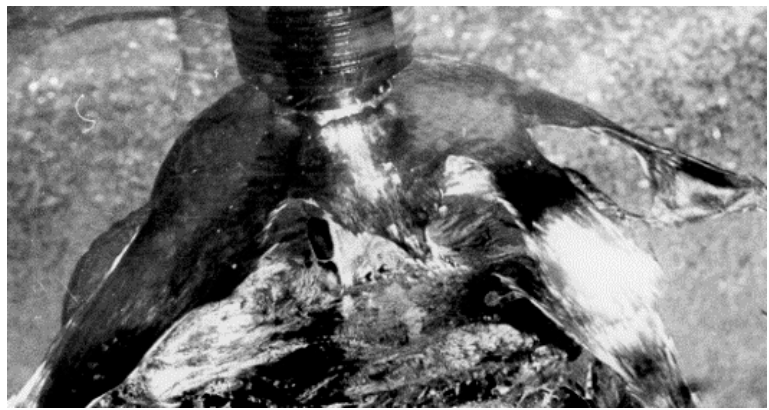

Fig. 11. Gallium film flow with oxidation of the melt in a weakly controlled nitrogen atmosphere

## 10. Film flow granulation technologies

Granulation technologies have specific applications, one of which is a production of the new materials from the uniform particles having the amorphous structure. Powder metallurgy has its own specific application, where the uniformity of small particles and their structure are not among the main requirements. By a development of the new materials with a structure close to the amorphous one, the main task is a cheap production of the uniform amorphous particles of the required size or with a narrow distribution by sizes (e.g. deviation from an average size 50% or so). The amorphous structure can be got by high cooling rates, therefore this task is as follows: cheap production of the nearly the same particles (granules) from the liquid metal, cooled (solidified) with the extremely high rate (we got $10^{4}$ $^0$C/s). Thus, the first one, in turn, requires starting this problem's solution from solving a task of obtaining the uniform drops from liquid metal in a comparably cheap way. We have solved it by a development of the theory and methods of the parametrically controlled disintegration of film flows.

In the case when the granules of the size over 2 mm are needed, the best are the jet type granulators [49-51, 54], which produce spherical particles with a deviation in sizes less than 5%. And the price of such granules' production is nearly the same as for example in the centrifugal granulation machines, where the size distribution of particles is wide (over 10 times in both sides from the required size). The traditional methods like the centrifugal ones fit well for production of the diverse metal granules including the high-temperature ones (e.g. steel) without the strict requirements to a size distribution and to a uniformity of cooling. The ultrasound methods allow getting the small particles, but they are much more expensive and have much lower productivity.

## 11. The conclusions and discussion

The developed granulation technology based on the controlled film flow phenomena allow obtaining the unique amorphous and special particles (e.g. with internal filling oriented in an electromagnetic field) for a production of the super hard metals, as well as the materials with the functionally unique properties. The known granulation machines (e.g. centrifugal) produce a wide spectrum of particle sizes and have much below cooling rate so that cannot compete with our technology. But we cannot produce granules from the high-temperature metals (steel) yet.